# Quantum search followed by classical search versus quantum search alone

P. R. M. Sousa,   F. V. Mendes   and   R. V. Ramos

pauloregisms@gmail.com   fernandovm@gmail.com   rubens.viana@pq.cnpq.br

*Lab. of Quantum Information Technology, Department of Teleinformatic Engineering – Federal University of Ceara - DETI/UFC, C.P. 6007 – Campus do Pici - 60455-970 Fortaleza-Ce, Brazil.*

In this work, we show that the usage of a quantum gate that gives extra information about the solution searched permits to improve the performance of the search algorithm by switching from quantum to classical search in the appropriated moment. A comparison to the case where only quantum search is used is also realized.

## 1. Introduction

Grover's quantum search [1] is a very useful and elegant quantum algorithm that has been extensively studied [2-5]. It has a quadratic speed up and, hence, it can find the searched element in a database with $N$ elements making only $O(N^{1/2})$ queries while the best classical algorithm requires $O(N)$ queries. In this work, we discuss the usage of quantum search followed by classical search and compare its performance to the case where only quantum search is used.

## 2. Quantum and classical search working together

Let us start by considering the following problem: One is looking for the bit string $x_{sol}$ that satisfies $f(x_{sol}) = y$. In order to solve this problem using quantum search, we assume the following quantum gate, which depends on the solution $x_{sol}$, exist: $U_1|a\rangle|0\rangle = |a\rangle|D_H(a,x_{sol})\rangle$, where $D_H(a,x_{sol})$ is the Hamming distance between the bit strings $a$ and $x_{sol}$. Obviously, $U|a\rangle|0\rangle = |a\rangle|0\rangle$ implies $a = x_{sol}$. Hence, the Grover's quantum algorithm can be used to find the solution $x_{sol}$ by using an oracle that recognizes the bit string $|0\rangle$ at the second register. It requires $\lceil(\pi/4)\sqrt{2^n}\rceil$ oracle calls, where $n$ is the number of bits of $x_{sol}$. Our proposal for solving the same problem is as follows: Firstly, the quantum search is used. In this case, the quantum states recognized by the oracle as solutions of the quantum search are those bit sequences having at most $k$ bits different from $x_{sol}$, that is, $D_H(a,x_{sol}) \leq k$. Thus, the output of the quantum search is a superposition of all bit sequences with Hamming distance (with respect to $x_{sol}$) equal or lower than $k$. In this case, the number of marked states is $M = \sum_{i=0}^{k}\binom{n}{i}$ and the number of oracle's calls in the quantum search is $N_G = (\pi/4)\sqrt{2^n/M}$. A measurement is realized and an $n$-bit sequence is obtained. At this moment, one knows this bit sequence is different from $x_{sol}$ in at most $k$ bits and a classical search can be used to find $x$ in this new database with $M$ elements. This will take in average $N_C = M/2$ queries. Hence, the total number of queries of the quantum-classical algorithm, $N_{GC} = N_G + N_C$, and its gain compared to the quantum search alone, $G = N_{GC}/\left[(\pi/4)\sqrt{2^n}\right]$, are, respectively

$$N_{GC} = \frac{\pi}{4}\sqrt{2^n / \sum_{i=0}^{k}\binom{n}{i}} + \frac{1}{2}\sum_{i=0}^{k}\binom{n}{i} \tag{1}$$

$$G = \left[\sum_{i=0}^{k}\binom{n}{i}\right]^{-1/2} + \frac{2}{\pi}\left(\sum_{i=0}^{k}\binom{n}{i}\bigg/2^{n/2}\right). \tag{2}$$

Observing (2), one sees that $k = 0$ recovers the pure quantum search while $k = n$ recovers the pure classical search. There is an optimal value for $k = k_{opt}$ that minimizes $G$. Furthermore, when $n$ grows, $k_{opt}$ also grows and $G$ tends to zero, showing the number of queries of the 'quantum+classical' algorithm is much smaller than the number of queries required by the quantum search working alone. Some examples of values for $n$, $k$ and $G$ are shown in Table 1.

| N | k | G | N | k | G |
|---|---|---|---|---|---|
| 100 | 6 | $1.586 \times 10^{-5}$ | 600 | 36 | $1.817 \times 10^{-30}$ |
| 200 | 12 | $1.536 \times 10^{-10}$ | 700 | 43 | $1.646 \times 10^{-35}$ |
| 300 | 18 | $1.531 \times 10^{-15}$ | 800 | 49 | $1.348 \times 10^{-40}$ |
| 400 | 24 | $1.576 \times 10^{-20}$ | 900 | 55 | $1.175 \times 10^{-45}$ |
| 500 | 30 | $1.67 \times 10^{-25}$ | 1000 | 61 | $1.094 \times 10^{-50}$ |

Table 1 – $G$ versus $n$ and $k$

The problem just described is not interesting in practice because there is a smart classical algorithm able to solve it in $O(n)$: Flip the first bit, if the Hamming distance to the solution decreases after flipping, then the new value of the first bit is the correct value otherwise, the original value of the first bit remains. Repeat the same steps to the following bits till get Hamming distance equal to zero.

A more hard situation for the same problem is to assume that, instead of $U_1$ that calculates the Hamming distance, only the following quantum gate, which depends on the solution $x_{sol}$, is available: $U_2|a\rangle|0\rangle = |a\rangle|h_k(D_H(a,x_{sol}))\rangle$ where $h_k(D_H(a,x_{sol})) = 0$ if $D_H(a,x_{sol}) > k$ and $h_k(D_H(a,x_{sol})) = 1$ if $D_H(a,x_{sol}) \leq k$. Once more, the quantum states recognized by the oracle as solutions of the quantum search (bit 1 in the second register) are those bit sequences having at most $k$ bits different from $x_{sol}$ and, hence, equations (1) and (2) are still correct. On the other hand, the (pure) smart classical algorithm cannot be used because one does not know the Hamming distance to the solution, but only if it is lower than *k* or not.

A related but more complicated problem is as follows: One is looking for the bit string $x_{sol}$ that satisfies $f(x_{sol}) = y$. In order to solve this problem using quantum search, we assume the following quantum gate, which depends on the solution $x_{sol}$, exist: $U_3|a\rangle|0\rangle = |a\rangle|g(a,x_{sol})\rangle$, where $g(a,x_{sol})$ is a known function. Furthermore, it is known that $D_H(a_i,a_j) \leq k$ for all $a_i$ and $a_j$ obeying the conditions $g(a_i,x_{sol}) \leq l$ and $g(a_j,x_{sol}) \leq l$. For this case, one has the following total number of queries and gain

$$N_{GC} = \frac{\pi}{4}\sqrt{2^n/M(g,l)} + \frac{1}{2}\sum_{i=0}^{k}\binom{n}{i} \tag{3}$$

$$G = \left[M(g,l)\right]^{-1/2} + \frac{2}{\pi}\left(\sum_{i=0}^{k}\binom{n}{i}\bigg/2^{n/2}\right). \tag{4}$$

In (3)-(4) $M(g,l)$ is the number of $a$'s that obey the condition $g(a,x_{sol}) \leq l$. Obviously, one must have $M(g,l) \leq \frac{1}{2}\sum_{i=0}^{k}\binom{n}{i}$.

## 3. Conclusions

The usage of a quantum gate that gives extra information about the solution of a search problem permits to improve the performance of the search algorithm by switching from quantum to classical search in the appropriated moment.